\documentclass[final,3p,times]{elsarticle}
\usepackage{amssymb}
\usepackage{amsbsy}
\usepackage{amsmath}
\usepackage{color}
\usepackage{mymacros}
\usepackage{pdflscape}
\usepackage{afterpage}
\usepackage{capt-of}
\usepackage[normal,tight,center]{subfigure}
\usepackage{array}
\usepackage{calc}
\usepackage[thinlines]{easytable}
\usepackage{comment}
\usepackage{tabu}
\usepackage[algo2e,ruled,linesnumbered]{algorithm2e}
\usepackage{booktabs} % For formal tables
\usepackage{mathtools}

\newcommand{\F}[3] {{F_{#1}^{{#2},{#3}}}}

\journal{Journal of Computational Physics}

\begin{document}

\begin{frontmatter}



\title{Anomaly detection in scientific data using joint statistical moments}

\author[label1]{Konduri Aditya\corref{cor1}} %\fnref{fnt1}}
\ead{akondur@sandia.gov}
\cortext[cor1]{Corresponding author.}
\author[label1]{Hemanth Kolla}
\ead{hnkolla@sandia.gov}
\author[label1]{W. Philip Kegelmeyer}
\ead{wkp@sandia.gov}
\author[label2]{Timothy M. Shead}
\ead{tshead@sandia.gov}
\author[label3]{Julia Ling}
\ead{jling@citrine.io}
\author[label2]{Warren L. Davis IV}
\ead{wldavis@sandia.gov}
\address[label1]{Sandia National Laboratories, Livermore, CA 94550, United States}
\address[label2]{Sandia National Laboratories, Abluquerque, NM 87123, United States}
\address[label3]{Citrine Informatics, Redwood City, CA 94063, United States}

\begin{abstract}
We propose an anomaly detection method for multi-variate scientific data based on analysis of high-order joint moments. Using kurtosis as a reliable measure of outliers, we suggest that principal kurtosis vectors, by analogy to principal component analysis (PCA) vectors, signify the principal directions along which outliers appear. The inception of an anomaly, then, manifests as a change in the principal values and vectors of kurtosis. Obtaining the principal kurtosis vectors requires decomposing a fourth order joint cumulant tensor for which we use a simple, computationally less expensive approach that involves performing a singular value decomposition (SVD) over the matricized tensor. We demonstrate the efficacy of this approach on synthetic data, and develop an algorithm to identify the occurrence of a spatial and/or temporal anomalous event in scientific phenomena. The algorithm decomposes the data into several spatial sub-domains and time steps to identify regions with such events. Feature moment metrics, based on the alignments of the principal kurtosis vectors, are computed at each sub-domain and time step for all features to quantify their relative importance towards the overall kurtosis in the data.  Accordingly, spatial and temporal anomaly metrics for each sub-domain are proposed using the Hellinger distance of the feature moment metric distribution from a suitable nominal distribution.  We apply the algorithm to a one-dimensional turbulent auto-ignition combustion case and demonstrate that the anomaly metrics reliably capture the occurrence of auto-ignition in relevant spatial sub-domains at the right time steps. 
\end{abstract}

\begin{keyword}

{Anomaly detection, Scientific computing, Co-Kurtosis, Tensor decomposition, Hellinger distance, Auto-ignition }

\end{keyword}

\end{frontmatter}

\section{Introduction}

Anomaly detection is such a widely studied topic, and has found numerous applications in various contexts, that it defies easy generalization. Nonetheless, the vast majority of applications that have embraced anomaly detection methods have characteristics that may not be representative of scientific data.  Chandola {\it et al.} \cite{ChandolaBK2009} emphasize that the key aspects of anomaly detection include the nature of input data, type(s) of anomaly and output of anomaly detection. In all these aspects, scientific data have distinctly different attributes compared to all other domains. As the scale of scientific investigations keeps ever increasing, robust anomaly detection is becoming increasingly critical. One of the key findings of a Department of Energy Workshop on mathematics of data \cite{MAPD} is ``\emph{near real-time identification of anomalies in streaming and evolving data is
needed in order to detect and respond to phenomena that are either short-lived or urgent}''.

Some of the challenges of anomaly detection in scientific data stem from the following attributes:
\begin{itemize}
\item Multi-variate, multi-physics phenomena: the observations are of numerous variables (tens to hundreds) that represent coupled non-linear physics and hence elude easy assumptions about statistical (in)dependence.
\item Multi-scale dynamics: the different observed variables span vastly different orders of magnitude since they are active at different scales in space and time. 
\item The observed data are not discrete but rather continuous and smoothly varying over multiple decades. With computational power ever increasing, the numerical resolution of investigations, e.g. scientific computing, is increasingly finer over a broader range of scales.
\item In most cases the field variables are not Gaussian. Examples of non-Gaussian distributions include beta or bi-modal shape distributions for reacting scalars in turbulent combustion and log-normal shape for dissipation in turbulence.
\item We are focusing on data rich, not data-sparse, scenarios like extreme-scale computational simulations or extremely well resolved measurements/observations. 
\end{itemize}

There are many scenarios where detecting scientific anomalies may be critically important. 
Many scientific investigations involve large quantities of rapidly streaming data: massively parallel computational fluid dynamics (CFD) simulations, large-scale particle accelerator data, real-time climate and meteorological simulations, etc.
Identifying anomalies as they occur can help judicious steering of these investigations (e.g. trigger analyses, data check pointing, mesh/time-step refinement, model parameter refinements, etc.).
Across these vast and varied scientific domains a general definition of an ``anomaly'' may be elusive and even within one domain it could be problem or regime dependent. Yet physics-driven anomalies, as opposed to investigative/measurement anomalies, occur in these settings and are important to detect. A semi-supervised method based on for random forests has been investigated in \cite{ling2017}.

We clarify that our focus is on identifying anomalous events in scientific investigations, if and when they occur in streaming scientific data, and not whether a specific observation is anomalous relative to the rest. To this end,  we want to develop an unsupervised methodology for detecting \emph{statistically identifiable but anomalous scientific phenomena e.g. ignition events in turbulent combustion and cyclones in climate simulations.}
Our intuition is that these phenomena have a statistical signature measurable in the higher statistical joint moments. Accordingly, we propose a methodology that is centered on analyzing high-order joint moments in multi-variate scientific datasets.
%For the purpose of this study we restrict ourselves to a distributed HPC setting where the computational domain is distributed on a parallel machine and the governing equations are marched in time. 
The anomalous event may occur at any time on some subset(s) of the spatial domain.
%and may need to be detected globally.
%This would apply for any distributed processing of streaming scientific data, so the approach is more broadly applicable beyond just HPC. 

The central hypothesis of the approach we propose is as follows:
\begin{itemize}
\item In data rich settings like scientific computing, anomalies have a discernible statistical signature since the data samples are usually large in number. 
\item Higher statistical moments, in particular kurtosis, are good indicators of outliers. 
\item For multi-variate data, by analogy to principal component analysis (PCA), principal vectors of the fourth order joint moment tensor -- principal kurtosis vectors -- signify directions along which outliers lie.
\item The occurrence of a anomalous event manifests as a \emph{sudden detectable change in the principal kurtosis vectors}.
\end{itemize}
We also note that, in choosing to analyze statistical joint moments, we assume that the large data sizes ensure that higher moments are reasonably converged. Moreover, joint moments can be computed in a computationally efficient manner and fast, single-pass algorithms for computing arbitrary order joint moments over distributed datasets are readily available in literature \cite{PebayTKB2016}. From a scientific computing perspective these algorithms have a high compute intensity and have regular, contiguous memory access patterns. They are scalable and likely to be efficient compared to algorithms that may be based on, say, decision trees.

\section{Background}

An anomaly can be loosely defined as an occurrence of something that is ``abnormal", ``atypical" or ``unexpected" \cite{Grubbs1969}. This is predicated on the assumption that what may be normal/typical/expected is well known and well defined, which may not always be the case. Accordingly, a wide variety of methods across various domains have come to be considered as anomaly detection methods. We first present a brief summary of anomaly detection methods from recent review papers that survey the literature on this topic. We will then discuss the applicability of existing methods, or lack thereof, for our domain of interest, anomalous events in scientific investigations. 

\subsection{Previous work}

Recent review papers by Chandola {\it et al.} \cite{ChandolaBK2009}, Campos {\it et al.} \cite{CamposZSCMSAH2016} and Goldstein and Uchida \cite{GoldsteinU2016} provide a useful survey and overview of various anomaly detection methods. They all conclude that it may be difficult to generalise, and compare, the various methods which stems from the inherent difficulty in defining, across domains, reasonable measures of what may be deemed as ``normal" and ``anomalous". Nonetheless, a broad categorization may be discerned. The types of anomalies usually sought are broadly three: 
\begin{itemize}
\item {\it point anomalies}: whether a given individual sample or observation is anomalous,
\item {\it collective anomalies}: whether a collection of samples/observations when considered together is anomalous, even if individual samples are not,
\item {\it contextual anomalies}: whether a point or collection of samples is otherwise normal, but anomalous given a specific context.
\end{itemize}
The vast majority of existing methods deal with the first kind, {\it point anomalies}. Accordingly the methods result in the identification of an anomaly at the individual sample level. Far fewer methods are devoted specifically to {\it collective} and {\it contextual anomalies}, and sometimes these problems are transformed to or posed as a {\it point} anomaly detection problem. The methods can also be distinguished based on whether the result is an anomaly score or a binary label. Furthermore, the methods span the full spectrum of {\it supervised}, {\it semi-supervised} and {\it unsupervised} paradigm. While we will not attempt an overview here, suffice it to say that the most commonly used {\it unsupervised} methods (see \cite{CamposZSCMSAH2016,GoldsteinU2016}) are clustering or nearest-neighbour based and involve measures of distance (with respect to nearest neighbours, global/local clusters), or density, the intuition being outlier samples have a large distance from normal samples or occur in regions of low density.  The more comprehensive survey of Chandola {\it et al.} establishes other prominent categories such as classification-based, statistical (parametric and non-parametric), information theoretic and spectral anomaly detection methods.

The specific setting of interest to us, as detailed in the next subsection, is {\it unsupervised} detection of {\it anomalous events} in scientific data, which may be considered closest to the {\it collective} anomaly detection class of methods. Such methods have been developed in the context of sequential anomaly detection, spatial anomaly detection and graph anomaly detection \cite{ChandolaBK2009}. However, none of these methods appear to be readily applicable for detecting anomalous scientific events, as described below, motivating the need for a new method.

\subsection{Anomalous or extreme events in scientific data}

Our interest is specifically in identifying anomalous events in scientific data which are sometimes also referred to as extreme events. These events represent genuine scientific phenomena that may be rare and/or extreme but are not necessarily spurious, as the interpretation may be for {\it point} anomalies. Some examples are ignition/extinction events in combustion data, tornadoes in climate data, crack propagation in fracture mechanics, etc. In a scientific investigation such extreme events are, by definition, not indicated by any individual sample observation, but by a group of observations, rendering {\it point} anomaly detection methods ineffective. Of the {\it collective} anomaly detection methods described by Chandola {\it et al.} \cite{ChandolaBK2009}, graph based techniques are not obviously applicable since few scientific data sets are represented using graphs. Continuum mechanics (e.g. solid/fluid mechanics) scientific data, which are representative of a broad class of scientific applications, consider joint spatio-temporal domains (discretized spatial mesh and discrete time instances) and hence a purely sequence-based (time domain) or spatial-anomaly based techniques will not capture a joint spatio-temporal anomaly. This is the very setting for our work. 

%\textcolor{red}{TO DO: Finish typing out this line of discussion.}

For a vast majority of statistical learning applications the assumption of normally distributed data is reasonable and hence all the statistical information is encapsulated in the covariance matrix. However, for scientific data the distributions are not often Gaussian and there is relevant statistical information in moments higher than the second (variance). Furthermore, scientific data often represent tightly coupled physical processes and hence the observed variables are, more often than not, statistically dependent.  Both aspects suggest that joint distributions, and joint moments higher than second order (covariance), are relevant, as opposed to marginal distributions and moments. 
%the co-variance, are relevant, as opposed to marginal distributions and moments. 

For the specific case of anomaly detection, then, it is a matter of analyzing the appropriate higher joint moment. By definition, the outlier samples have an increasingly greater contribution the higher the moment. However, from a practical perspective, we suggest that the fourth moment -- kurtosis -- is appropriate. While kurtosis has been often held as a measure of ``peakiness'' or ``flatness'', we refer to the paper by Westfall \cite{Westfall2014} that establishes kurtosis as an unambiguous measure of ``either existing outliers (for the sample kurtosis) or propensity to produce outliers (for the kurtosis of a probability distribution).'' Westfall illustrates \cite{Westfall2014} that as kurtosis increases, the contribution to it from the portion of data centered around the mean becomes vanishingly small, no matter how the center is defined (i.e. no matter how many multiples of standard deviation). In the limit of infinite kurtosis the contribution from any finite portion centered around the mean is zero no matter how large this portion. These results suggest that kurtosis is a reliable enough measure of outliers. Accordingly, our approach is centered around analyzing the joint fourth moments -- the co-kurtosis -- which is a fourth order tensor (co-variance is a second order tensor i.e. a matrix). To reiterate, \emph{we seek, by analogy to PCA, principal vectors of the co-kurtosis tensor which can be interpreted as principal directions along which outliers lie.} This effectively becomes a symmetric tensor decomposition problem. 

%\subsection{Statistical joint moments}
\subsection{Joint moment tensor decomposition}
Joint moment tensors have been analyzed in different settings, but are usually considered expensive due to the curse of dimensionality e.g. the size of the fourth moment tensor is $N_f^4$, where $N_f$ is the number of random variables or features. By definition, the tensor is symmetric which means that the number of unique entries is smaller than $N_f^4$, but to leading order the scaling is still a fourth power of $N_f$. Jondeau {\it et al.}, \cite{JondeauJR2015}, recognizing that non-normality is important in analyzing financial market data, consider decompositions of the co-skewness and co-kurtosis tensors for analysis of stock market data, and conclude that the first few factors of the decomposition of these tensors contain useful information about market returns. The fourth moment tensor is also at the heart of mathematical underpinnings of Independent Component Analysis (ICA), which we will review shortly. 

Since most of the properties of matrix factorizations do not extend in general to higher order tensors \cite{KoldaB2009}, the appropriate decomposition technique, and its associated interpretation, needs to be chosen carefully. We briefly present an overview of major classes of tensor decompositions, that may be applicable to the present problem, before explaining the particular method chosen in this study. We use third order tensors, only for illustration purposes, in the following discussion.

\subsubsection{Symmetric CP decomposition}

\begin{figure}[htb]
  \centering
  \includegraphics[trim=3cm 12cm 3cm 11cm,clip,width=9cm]{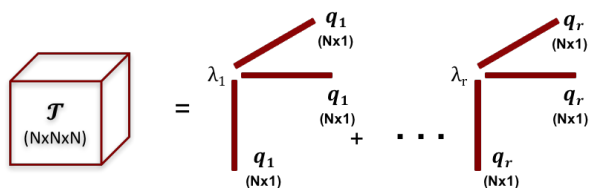}
  \caption{Symmetric CP decomposition of a third order symmetric tensor $\mathcal{T}$.}
  \label{f:sym_CP_decomp}
\end{figure}

Canonical polyadic (CP) decomposition seeks a factorization of a tensor as a sum of outer products of real-valued vectors. Mathematically, for a third order tensor $\mathcal{T}$, this can be expressed as
\begin{equation}
\mathcal{T} = \sum_{i=1}^r \lambda_i ~ x_i \otimes y_i \otimes z_i,
\end{equation}
where $\otimes$ denotes the outer product between the sets of vectors $x_i, y_i, z_i$ and the number of such sets sought, $r$, is the rank of the decomposition. Even when the tensor $\mathcal{T}$ is symmetric, it is possible to seek a decomposition which is asymmetric, that is, $x_i$, $y_i$ and $z_i$ are not equal. However, a symmetric decomposition always exists for a symmetric tensor \cite{ComonGLM2008} such that  
\begin{equation}
\mathcal{T} = \sum_{i=1}^r \lambda_i ~ x_i \otimes x_i \otimes x_i,
\end{equation}
as illustrated in Fig. \ref{f:sym_CP_decomp}, and $r$ in this case is referred to as the \emph{symmetric rank}.
%%%%%%%%
%%%%%%%%
While seemingly analogous to eigenvalue decomposition of a symmetric matrix (PCA can be interpreted as the eigenvalue decomposition of the co-variance matrix), a few key differences remain. The rank $r$ in the above decomposition is not a unique number, but lies within certain bounds \cite{ComonGLM2008}. Moreover, it is not necessary that the decomposition be orthogonal, which is the case for eigenvalue decomposition of a matrix. Orthogonal decompositions of symmetric tensor do not exist in general \cite{Kolda2015}, but if it is known to exist for a specific tensor the decomposition problem can be reduced to a matrix factorization problem \cite{Kolda2015,AnandkumarGHKT2014}.  

\subsubsection{Higher order Singular Value Decomposition (HOSVD)}

\begin{figure}[htb]
  \centering
  \includegraphics[trim=5cm 11cm 5cm 11cm,clip,width=9cm]{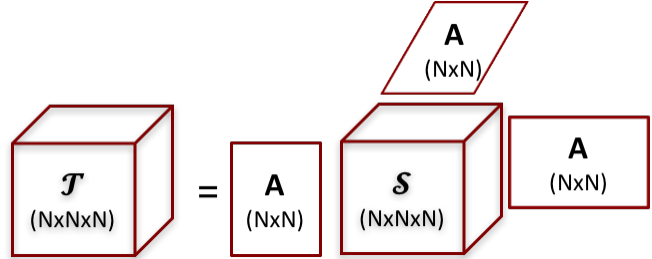}
  \caption{HOSVD of a third order symmetric tensor $\mathcal{T}$.}
  \label{f:sym_HOSVD}
\end{figure}

An alternate decomposition, which extends the matrix singular value decomposition (SVD) concept to higher order tensors is Higher-Order Singular Value Decomposition (HOSVD) \cite{DeLathauwerDMV2000}. For a general third order tensor, this is written mathematically as
\begin{equation}
\mathcal{T} = \mathcal{S} \times_1 {\rm U}^{(1)} \times_2 {\rm U}^{(2)} \times_3 {\rm U}^{(3)}
\end{equation}
where the core tensor, $\mathcal{S}$, has the same dimensions as $\mathcal{T}$. The factor matrices ${\rm U}^{(1)}$, ${\rm U}^{(2)}$ and ${\rm U}^{(3)}$ are orthogonal matrices that result from ``unfolding'' (matricizing)  $\mathcal{T}$ along modes 1, 2 and 3, respectively, and performing SVD over the resulting matrix. The symbol $\times_k$ denotes a ``mode-k'' tensor-matrix product (see \cite{DeLathauwerDMV2000} for details). For a symmetric tensor the result of ``unfolding'' is the same along any mode and hence the factor matrices become identical. This special case is illustrated in Fig. \ref{f:sym_HOSVD}. For more details on the mathematical properties and conventions of HOSVD, the analogies to and differences with matrix SVD, the reader is referred to De Lathauwer {\it et al.} \cite{DeLathauwerDMV2000}. The column vectors of the (identical) factor matrices may be interpreted by analogy to principal vectors in PCA. However, one key departure between HOSVD and matrix SVD is that the core tensor $\mathcal{S}$ is not purely diagonal but dense. This makes the interpretation of the column vectors of factor matrices, by analogy, not straightforward. 
%%%%%%%%
%%%%%%%%

\subsection{Independent Component Analysis}

In the present study the most useful approach for the fourth moment tensor decomposition appeared to be by way of analogy to Independent Component Analysis (ICA). ICA specifically deals with non-Gaussian random variables and considers a scenario where a set of statistically independent non-Gaussian random variables ${\rm s} = [s_1 ~s_2 ~...~s_q]^T$ are linearly mixed, superimposed with independent white noise, and observed as the set of random variables ${\rm y} = [y_1 ~y_2 ~...~y_p]^T$. Mathematically,
\begin{equation}
y = As + n
\label{eq:ICA_model}
\end{equation}
where $A \in \mathbb{R}^{p\times q}$ is the mixing matrix and $n \in \mathbb{R}^p$ is independent white noise, and ICA aims to identify the source set $\rm s$ given the observed set $\rm y$. 

The connection between ICA and higher moment tensor decomposition is explained in many papers (see De Lathauwer and Moore \cite{DeLathauwerM2001} or Anandkumar {\it et al.} \cite{AnandkumarGHKT2014}). Formally, given the statistical model in Eq. \ref{eq:ICA_model}, the fourth order cumulant tensor of $y$ (observed variables) is related to the outer product of column vectors, $a_i$, of the mixing matrix $A$ and the excess kurtosis, $\kappa_i$, of the individual sources $s_i$ \cite{Comon2002}:
\begin{equation}
C_4^y = \sum_{i=1}^q \kappa_i ~~a_i \otimes a_i \otimes a_i \otimes a_i.
\label{eq:ICA_tensor_decomp}
\end{equation}
The cumulant tensor is related to the fourth and second moment tensors as
\begin{eqnarray}
\left[C_4^y \right]_{i_1 i_2 i_3 i_4} = \mathbb{E}[y \otimes y \otimes y \otimes y ] - \mathbb{E}[y_{i_1} y_{i_2}] \mathbb{E}[y_{i_3} y_{i_4}] \nonumber \\
- \mathbb{E}[y_{i_1} y_{i_3}] \mathbb{E}[y_{i_2} y_{i_4}] - \mathbb{E}[y_{i_1} y_{i_4}] \mathbb{E}[y_{i_2} y_{i_3}],~~1 \leq i_1...i_4 \leq k
\label{eq:cumulant_tensor_def}
\end{eqnarray}
where $\mathbb{E}$ is the expectation operator.
These authors have also proposed algorithms for performing the tensor decomposition in the form of Eq. \ref{eq:ICA_tensor_decomp}. De Lathauwer and Moore \cite{DeLathauwerM2001} propose a method based on ``simultaneous third-order tensor diagonalization'', Anandkumar {\it et al.} \cite{AnandkumarGHKT2014} propose a robust variant of a tensor power method, while Kolda \cite{Kolda2015} proposes a method involving eigen value decomposition of a matrix that is linear combination of slices of the tensor. 

However, De Lathauwer and Moore \cite{DeLathauwerM2001} and Anandkumar {\it et al.} \cite{AnandkumarGHKT2014} also observe the connections to a much simpler matrix SVD problem. Specifically, in a discussion on computational complexity, Anandkumar {\it et al.} \cite{AnandkumarGHKT2014} note that if the cumulant tensor $C_4^y$ is unfolded into a matrix $M^y$ (the unfolding is invariant to choice of mode since $C_4^y$ is symmetric), then Eq. \ref{eq:ICA_tensor_decomp} can be transformed to
\begin{equation}
mat(C_4^y) \coloneqq M^y =  \sum_{i=1}^q \kappa_i ~~a_i \otimes vec(a_i \otimes a_i \otimes a_i)
\end{equation}
and hence the vectors $a_i$ can be determined from an SVD of $M^y$ ($mat$ and $vec$ denote operations that convert a tensor to a matrix and a vector, respectively). For its simplicity, ease of interpretation and computational cost considerations, we adopt this approach to decomposing the fourth moment tensor in this study. 

\subsection{Tests with synthetic data}
%Section where we demonstrate the tensor decomposition on toy data sets. Stress on the point that, depending on the nature of the data, PCA and kurtosis vectors can be different. 

The method chosen to identify principal directions of joint moment tensor has three simple steps: (a) construct the joint fourth cumulant tensor ($C_4^y$ in Eq. \ref{eq:cumulant_tensor_def}), (b) matricize the tensor ($M^y$) along any of the modes, and (c) perform SVD of the matrix $M^y$. The resulting singular vectors of the SVD (whether left or right singular vectors depends on how the tensor is matricized) are the desired principal vectors. We present tests of this method on synthetic datasets. All tests were conducted in Matlab and the tensor operations were performed using tensor toolbox \cite{TTB_Software}. Two bi-variate datasets are chosen and the main purpose is to illustrate that:
\begin{itemize}
\item the method recovers the principal vectors for datasets when they are known, by construction,
\item compare the principal kurtosis vectors with PCA vectors to show that they need not be equal even for simple datasets. 
\end{itemize}

 %%%%%%%%
\begin{figure}[htb]
  \centering
  \includegraphics[trim=1.5cm 7.5cm 3.5cm 8cm,clip,width=8cm]{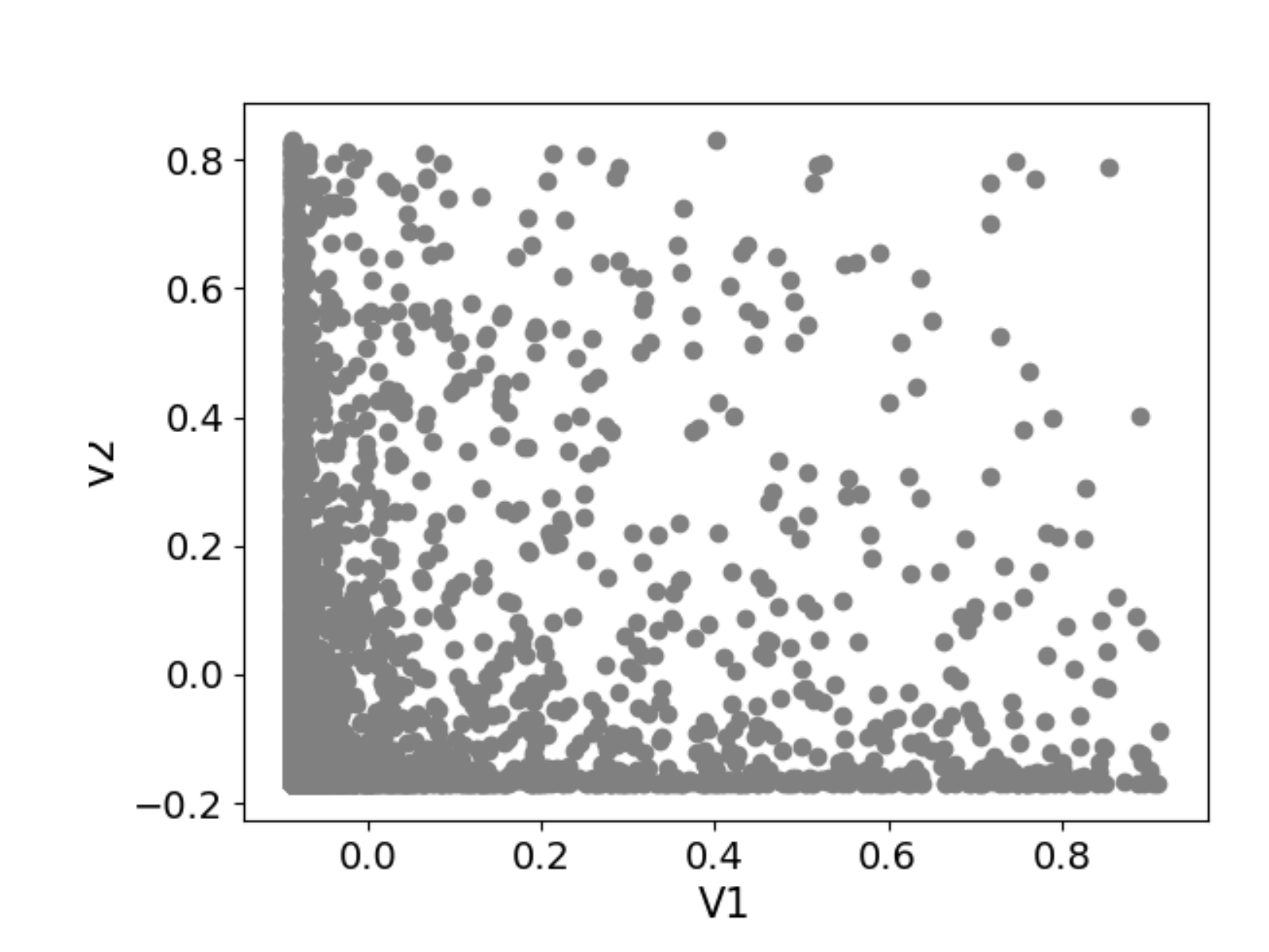}
  \includegraphics[trim=1.5cm 7.5cm 3.5cm 8cm,clip,width=8cm]{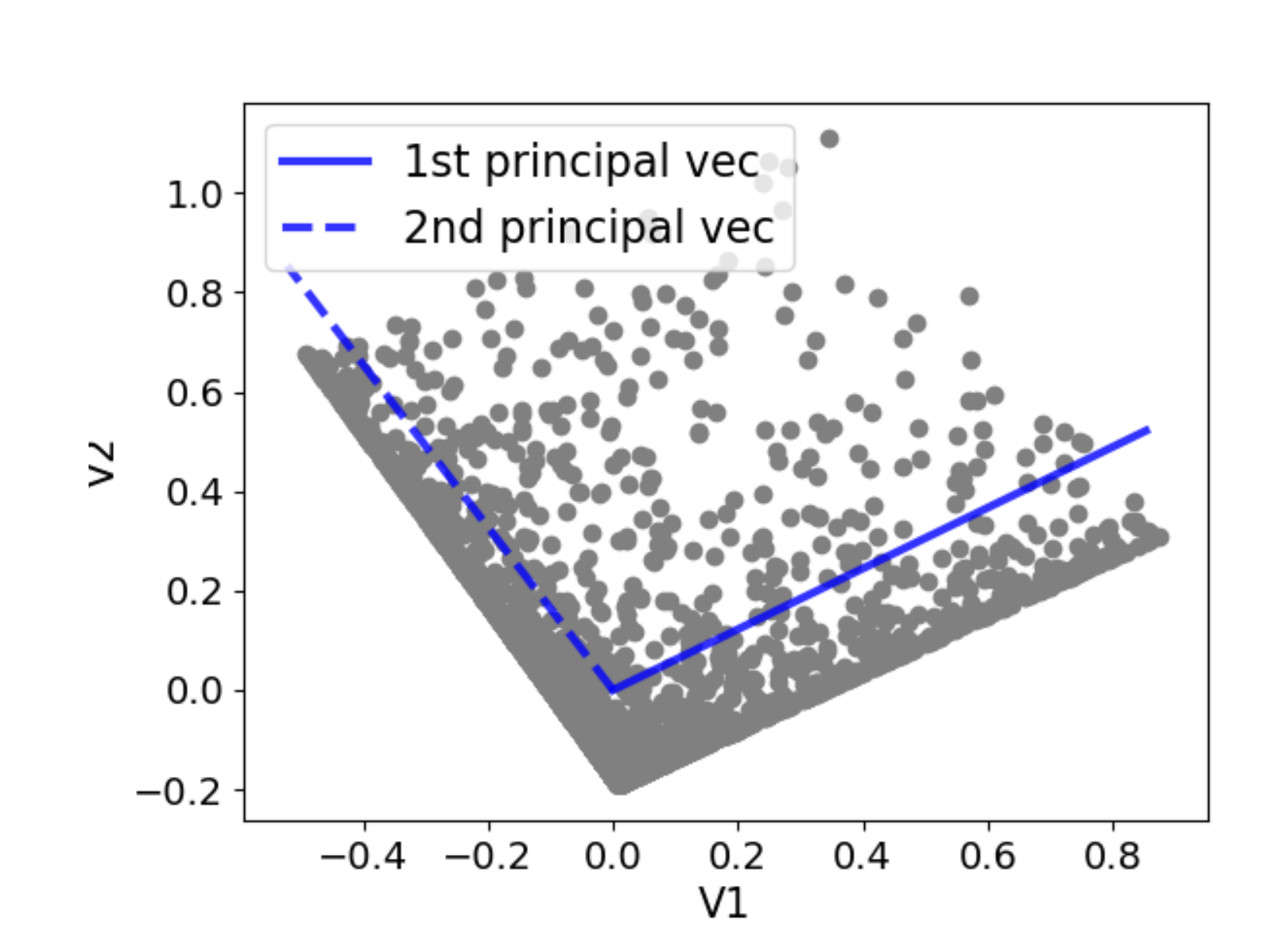}
  \caption{(Left) Synthetic bi-variate statistically independent dataset with beta marginal distributions. (Right) The independent bi-variate is ``mixed'' (i.e. rotated by $30^o$ to transform to a dataset with non-zero higher joint moments. The first (solid blue line) and second (dashed blue line) principal kurtosis vectors of this dataset are also shown.}
  \label{f:syn_beta_beta}
\end{figure}
%%%%%%%%

\begin{figure}[htb]
  \centering
  \includegraphics[trim=1.5cm 7.5cm 3.5cm 8cm,clip,width=8cm]{beta_beta_kurtvecs.pdf}
  \includegraphics[trim=1.5cm 7.5cm 3.5cm 8cm,clip,width=8cm]{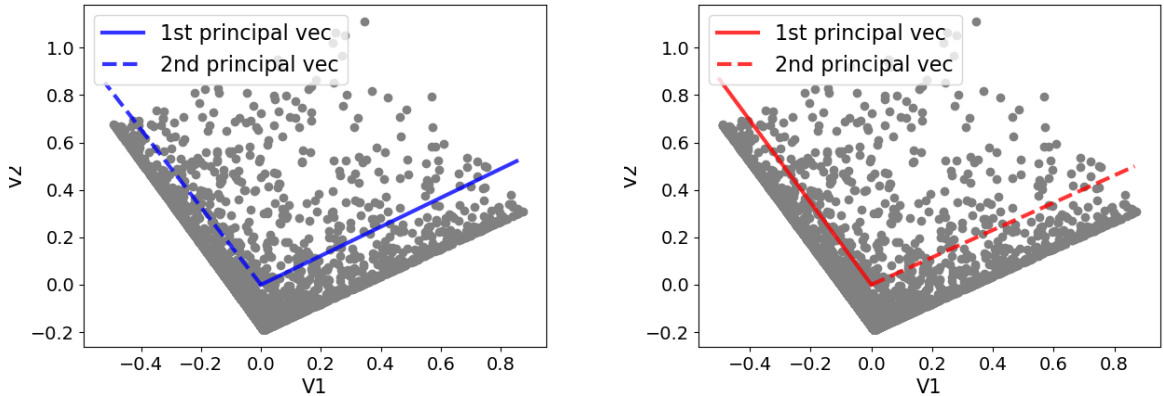}
  \caption{Comparison between the principal kurtosis vectors shown in blue (left) and the PCA vectors shown in red (right) for the synthetic dataset from Fig. \ref{f:syn_beta_beta}.}
  \label{f:syn_beta_beta_kurt_pca}
\end{figure}

For the first test a bi-variate dataset with known statistical joint moments is constructed in two steps. In the first step a bi-variate Gaussian copula with zero cross-correlation is sampled to ensure a statistically independent dataset. The inverse beta cumulative distribution function is applied over each copula generated vector such that the resulting dataset has beta marginal distributions, with known moments, and all the joint moments are approximately zero within sampling error. More importantly, the dataset is non-Gaussian and has non-zero higher order moments. The resulting dataset is shown in the left panel of Fig. \ref{f:syn_beta_beta}. The [$\alpha$, $\beta$] parameters of the beta distributions are chosen to be [1.0, 0.1] for the first variable $V_1$ (x-axis) and [1.0, 0.2] for the second variable $V_2$ (y-axis). These parameters were chosen carefully such that the excess kurtosis of $V_1$ is greater than that of $V_2$, whereas the variance of $V_2$ is greater than that of $V_1$. In the second step this independent dataset is ``mixed'' by performing an Euler rotation, and the resulting dataset now has non-zero joint moments. The first and second principal kurtosis vectors for this dataset are extracted using the described methodology and are verified to be equal to the vectors of the rotation matrix. These are shown in blue in the right panel of Fig. \ref{f:syn_beta_beta}. Crucially, the parameters of this dataset were so chosen such that the principal directions of variance (PCA vectors) are different from that of the kurtosis i.e. the first PCA vector and the first kurtosis vector are orthogonal to each other, as are the second. This is illustrated in Fig. \ref{f:syn_beta_beta_kurt_pca}.   

%%%%%%%%
%%%%%%%%

%%%%%%%%
\begin{figure}[htb]
  \centering
  \includegraphics[width=6cm]{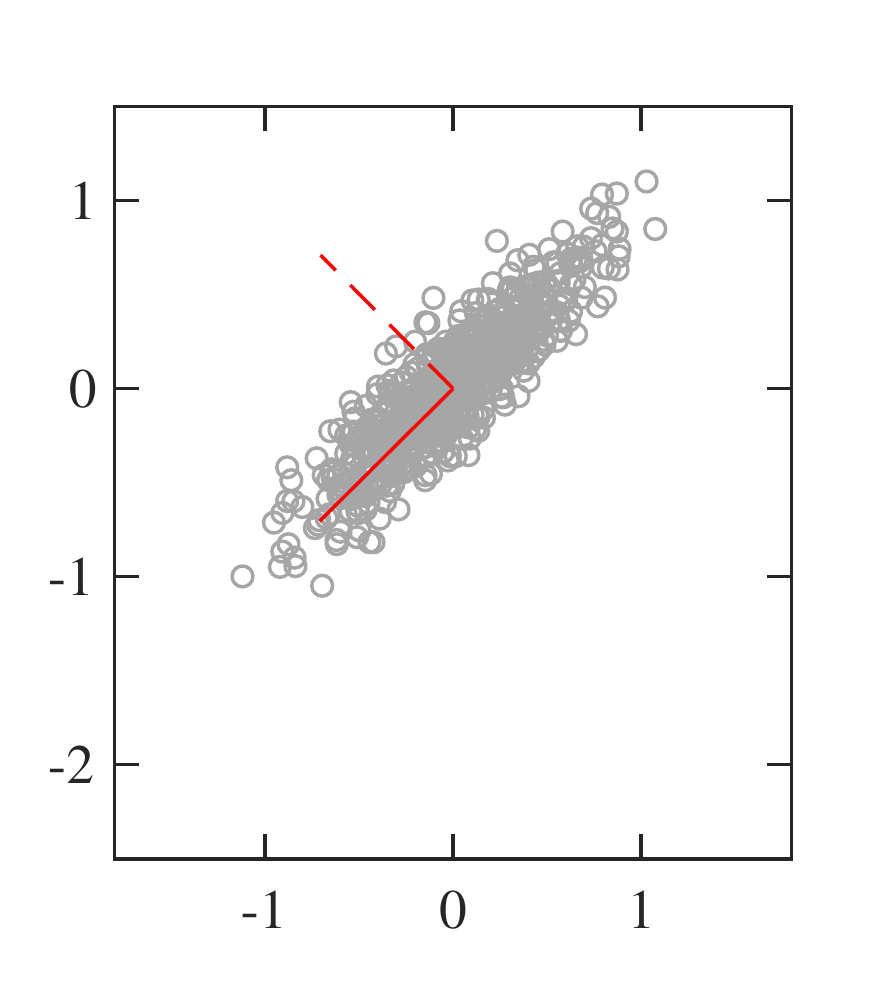}
  \bp
      \put(-90,0){$V_1$}
      \put(-175,95){\rotatebox{90}{$V_2$}}
  \ep
  \hspace{0.6cm}
  \includegraphics[width=6cm]{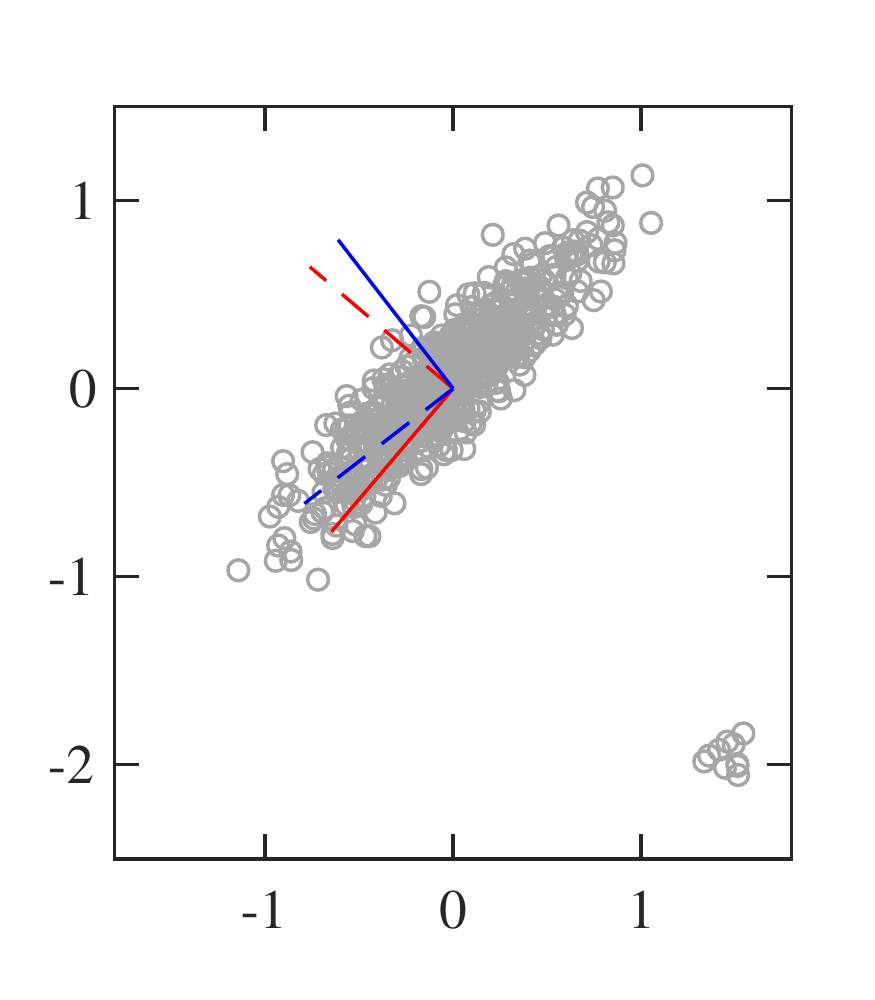}
    \bp
      \put(-90,0){$V_1$}
      \put(-175,95){\rotatebox{90}{$V_2$}}
  \ep
  \caption{(Left) An uncorrelated bi-variate Gaussian dataset with zero means and variances [1.0, 0.25] is mixed by rotation. The first (solid red) and second (dashed red) PCA vectors are shown. (Right) Gaussian dataset with a few samples converted to outliers. The principal kurtosis vectors (blue) are shown along with the PCA vectors (red).}
  \label{f:syn_data}
\end{figure}
%%%%%%%%

For the second test a dataset with a more visually obvious set of outliers is constructed. An uncorrelated bi-variate Gaussian dataset, with zero means and variances [1.0, 0.25] is generated, and then mixed by rotation. The left panel in Fig. \ref{f:syn_data} shows the dataset, along with the PCA vectors shown in red. Since the data has Guassian structure, the higher moments and their principal directions are ill-defined. Then, a small subset of samples in this dataset are chosen at random and placed far away from the rest, to mimic outliers, as shown on the right panel of Fig. \ref{f:syn_data}. The PCA vectors (red) and principal kurtosis vectors (blue) for the data with outliers are also shown. In comparison to the left panel, it is evident that the PCA vectors remain nearly unchanged. However, the principal kurtosis vectors are different from the PCA vectors, and clearly align with the direction the outliers lie. The results from both test datasets illustrate that principal kurtosis vectors need not be the same as PCA vectors, and are better at identifying directions of outliers.

\section{Anomaly detection algorithm}
In the previous section, we have shown that the variance or the kurtosis in data can be characterized in terms of the principal values and vectors of the joint moment tensors.
Anomalies, as they occur, result in a change in the distribution of the data reflecting a change in the magnitude of the principal values and the orientation of the principal vectors. We base our anomaly detection algorithm for smoothly varying scientific data on this concept, as explained in this section.

As mentioned earlier, anomalous events in scientific phenomena can appear locally in space and/or time. We choose to decompose the data into several spatial sub-domains and time steps to explicitly flag the locality of these events. 
For distributed streaming data, e.g. data from massively parallel simulations, such a decomposition is inherently present by way of domain decomposition. Let $N_d$ and $N_t$ be the number of spatial sub-domains and the number of time steps, respectively. In regard to the feature space, depending on the nature of anomaly and guidance from the scientific domain, not all the features may be relevant to identify its occurrence. Let $N_f$ be the number of relevant features to be used in the anomaly detection algorithm. 
The first step in the algorithm involves a data preprocessing stage, where we scale the data from each feature by subtracting its mean and dividing by the absolute spatial maximum.
It is quite common in multi-scale simulations that value ranges of different features are decades apart. Hence, scaling the data
%by subtracting the mean and dividing by the absolute maximum of each feature
ensures an equitable contribution to the joint moments from all the features.

Once the data is preprocessed, for each sub-domain $j$ at a given time step $n$, the joint moment tensor $\mathcal{T}^{j,n}$ is constructed. This symmetric tensor is decomposed using the two-step (matricize and perform SVD) method described in section 2.3, to obtain the principal values $\lambda_j$ and the principal vectors $\hat{v}_j$.
For a given phenomena, in the absence of anomalous events, the principal values and vectors would remain nearly the same in all the sub-domains. As mentioned above, the occurrence of an anomalous event will result in a significant change in the magnitude of the principal values and the orientation of the principal vectors. We, now, define a feature moment metric $\F{i}{j}{n}$ for each feature $i$ in a given sub-domain $j$ and time step $n$, which can be used to quantify the changes in the principal values and vectors. 
\begin{equation}
\F{i}{j}{n} = \frac{\sum\limits_{k=1}^{N_f} \lambda_k \left(\hat{e}_i \cdot \hat{v}_k\right)^2}{\sum\limits_{k=1}^{N_f} \lambda_k}
\end{equation}
It should be noted that $\hat{e}_i \cdot \hat{v}_k$ is effectively the $i$-th entry in the $k$-th vector $\hat{v}_k$. 
Since the set of vectors $\hat{v}_k$ are all unit vectors, by construction, the set of feature moment metrics in every spatial ($j$) and temporal ($n$) sub-domain sum to unity, i.e. $\sum\nolimits_{i=1}^{N_f} \F{i}{j}{n} = 1,~  \forall ~ j, n$. Accordingly, the moment metric for a given $i$ can be interpreted as a measure of the fraction of the overall moment (variance or excess-kurtosis as case may be) contained in feature $i$, in other words a distribution of the moment in the feature space. This distribution interpretation allows comparing the feature moment metrics between different sub-domains in space and different steps in time, and flag the occurrence of anomalous events in space and/or time.

If, as hypothesized, the statistical signature of anomalies is such that the distribution of feature moment metrics measurably changes, then distribution divergence metrics, such as $f$-divergence, can be used to quantify the change. We use the Hellinger distance, a symmetric measure of difference between two discrete distributions $P$ and $Q$:
\begin{equation}
D_{PQ} = \frac{1}{\sqrt{2}} \sqrt{ \sum_i  (\sqrt{p_i} - \sqrt{q_i})^2}  ~.
\end{equation}
The Hellinger distance lies between 0 and 1, and for a discrete distribution the distance is 1 when the two distributions being compared are exact complements of each other i.e. if $\forall~i$ when $p_i \neq 0$ and $q_i = 0$, and vice--versa.
Intuitively, for the anomaly containing spatial/temporal sub-domain, the Hellinger distance of the $p_i \equiv \F{i}{j}{n}$ from a nominal distribution $q_i$ would be large, and a suitable threshold of this distance can be used as an anomaly metric. The nominal set, $q_i$, can be chosen to be the spatial average such that the distance quantifies a spatial anomaly (at every time instance), whereas, setting $q_i$ to be the previous time distribution quantifies a temporal anomaly (in each spatial sub-domain). Accordingly, we define a spatial anomaly metric
\begin{equation}
M_1^n(j) =  \frac{1}{\sqrt{2}} \sqrt{ \sum_{i=1}^{N_f}  \left(\sqrt{\F{i}{j}{n} } - \sqrt{ \overline{F}_i^n } \right)^2},
\label{eq:M1}
\end{equation}
where $\overline{F}_i^n$ denotes the spatial (over $j$) average of $\F{i}{j}{n}$. A corresponding temporal anomaly metric can be defined as
\begin{equation}
M_2^j(n) =  \frac{1}{\sqrt{2}} \sqrt{ \sum_{i=1}^{N_f}  \left(\sqrt{\F{i}{j}{n} } - \sqrt{ F_i^{j,n-1} } \right)^2}.
\label{eq:M2}
\end{equation}
%DESCRIBE ANOMALY METRIC

%\subsection{Implementation}
We, now, proceed to describe the implementation of the anomaly detection algorithm, which is outline in Algorithm 1. The line 1 decomposes the data into different sub-domains and time steps. The relevant features for the algorithm are selected in line 2. Once these initialization steps are complete, the computations enter the time step loop in line 3. For each time step, the data is accessed with a sub-domain loop which begins in line 4. For a given time step and sub-domain, the data is scales in line 5 before computing the joint moment tensor in line 6. The tensor is then matricized in line 7 to perform SVD in line 8 to obtain the principal values $\lambda_j$ and principal vectors $\hat{v}_j$. The feature moment metrics and the anomaly metrics are computed in line 9 and 10, respectively. After these computations are performed over all the sub-domains, the anomaly metrics are compared with a threshold value in line 12 to flag the occurrence of any anomalous event.

\begin{algorithm2e}
\tcp{initialization}
$N_t$, $N_d$ $\leftarrow$ decompose data\; 
$N_f$ $\leftarrow$ select features\;
\tcp{time step loop}
\For{$n \leftarrow 1$ \KwTo $N_t$}{

\tcp{sub-domain loop}
\For{$j \leftarrow 1$ \KwTo $N_d$}{
scale data\;
$\mathcal{T}^{j,n}$ $\leftarrow$ compute joint moment tensor\;
matricize tensor $\mathcal{T}^{j,n}$ \;
$\lambda_j$, $v_j$ $\leftarrow$ perform SVD\;
$\F{i}{j}{n}$ $\leftarrow$ compute feature importance\;
$M_1^n(j)$, $M_2^j(n)$ $\leftarrow$ compute anomaly metrics\;
}
flag anomalous sub-domains;
}
\caption{Anomaly detection algorithm}
      \label{alg:ADA}
\end{algorithm2e}

\section{Autoignition test case}
Autoignition is a spontaneous combustion process where a small temperature rise due to exothermic reactions results in the self-ignition of a fuel in the presence of an oxidizer. This phenomenon plays a critical role in the operation of devices such as the compression ignition engines and gas turbines \cite{Dec09,Guethe09}. The characteristics of autoignition strongly depend on the fuel-oxidizer composition, thermodynamic and flow conditions, among others. Hence, several experimental and simulation studies are devoted to understand the autoignition phenomena at conditions relevant to practical devices.

As the chemical phenomena resulting in autoignition possess an exponential behavior, the inception of ignition kernels (parcels of burning mixtures) is very sensitive to the initial conditions. Also, the ignition events are short-lived in time. In practical devices, where spatial inhomogeneities persist, e.g. due to turbulence and mixing, these events are often highly localized in space. This makes it challenging to detect and analyze the phenomena \cite{Bennett2016}. In the combustion community, researchers often use simple threshold based metrics which are ad hoc in nature, or mathematically rigorous formulations such as chemical mode explosive analysis (CEMA) \cite{lu10} which are computationally expensive, to detect the autoignition events in complex turbulent reacting flows. 
%In the context of machine learning algorithms, the occurrence of ignition events can be labeled as anomalous either in time or in space, or in both space and time. 
We will use the anomaly detection algorithm described in section 3 to identify the occurrence of auto-ignition events in a representative combustion simulation. 

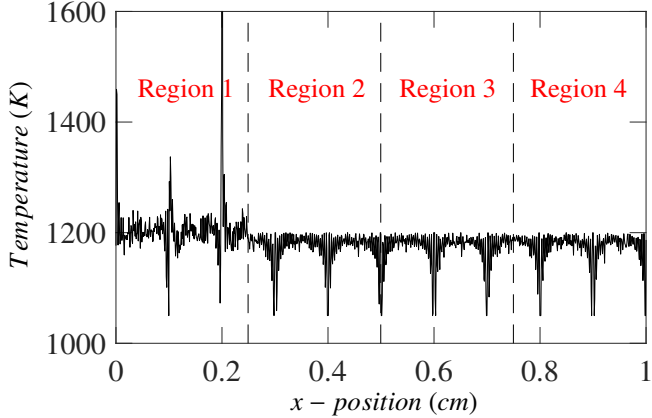
\begin{figure}[htb]
  \centering
  \includegraphics[width=9cm]{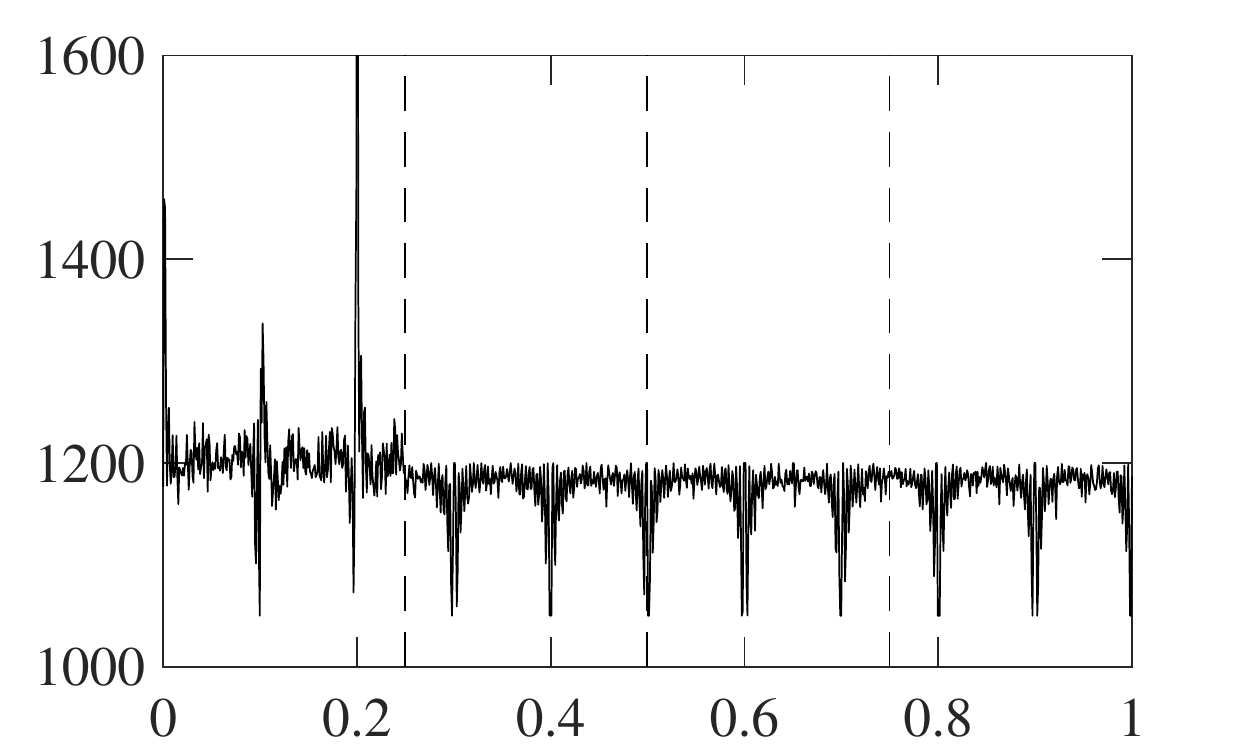}
      \bp
      \put(-160,-8){$x-position$ $(cm)$}
      \put(-265,45){\rotatebox{90}{$Temperature$ $(K)$}}
      \put(-217,110){\colr Region 1}
      \put(-168,110){\colr Region 2}
      \put(-119,110){\colr Region 3}
      \put(-70,110){\colr Region 4}
  \ep
  \caption{Profile of the temperature initial condition. Dashed lines represent the sub-domain boundaries of the decomposed spatial domain.}
  \label{f:temp_ic}
\end{figure}

A spatially one-dimensional time-varying simulation of an auto-ignitive mixture is performed using a direct numerical simulation solver, S3D \cite{CHEN+2009}, which solves the reacting compressible flow governing equations. The governing partial differential equations are discretized using explicit eight-order central difference schemes for spatial derivatives and six-stage fourth-order Runge-Kutta method for time integration. Explicit tenth-order filters are used to remove spurious numerical oscillations. A 12-species 29-reactions syngas mechanism \cite{hawkes2007scalar} is used for the combustion chemistry. The one-dimensional domain is 1cm long and is discretized using 1024 grid points. 
The domain is initialized with premixed fuel-oxidizer mixture comprised of species concentrations $0.6CO + 0.4H_2 + 0.5(O_2 + 3.76N_2)$. The initial pressure is set to 4 atm and the flow is quiescent. A spatially inhomogeneous temperature field with a mean of 1200K is imposed using a linear combination of sinusoidal waves with random phases. In addition, a spike in the temperature is superimposed in the first quarter of the domain, which ensures that this portion ignites first, resulting in a spatially anomalous ignition event. The initial temperature profile is shown in Fig. \ref{f:temp_ic}. A periodic boundary condition is imposed for solving the equations. The simulation is time advanced up to $20 \mu s$, using fixed  time-steps of $0.001 \mu s$. The data from the simulation is stored at equal intervals of $1 \mu s$, which is one-tenth of the ignition delay time based on a homogeneous mixture at a temperature of 1200K.
%%%%%%%%
%%%%%%%%

\begin{figure}[htb]
  \centering
  \includegraphics[width=9cm]{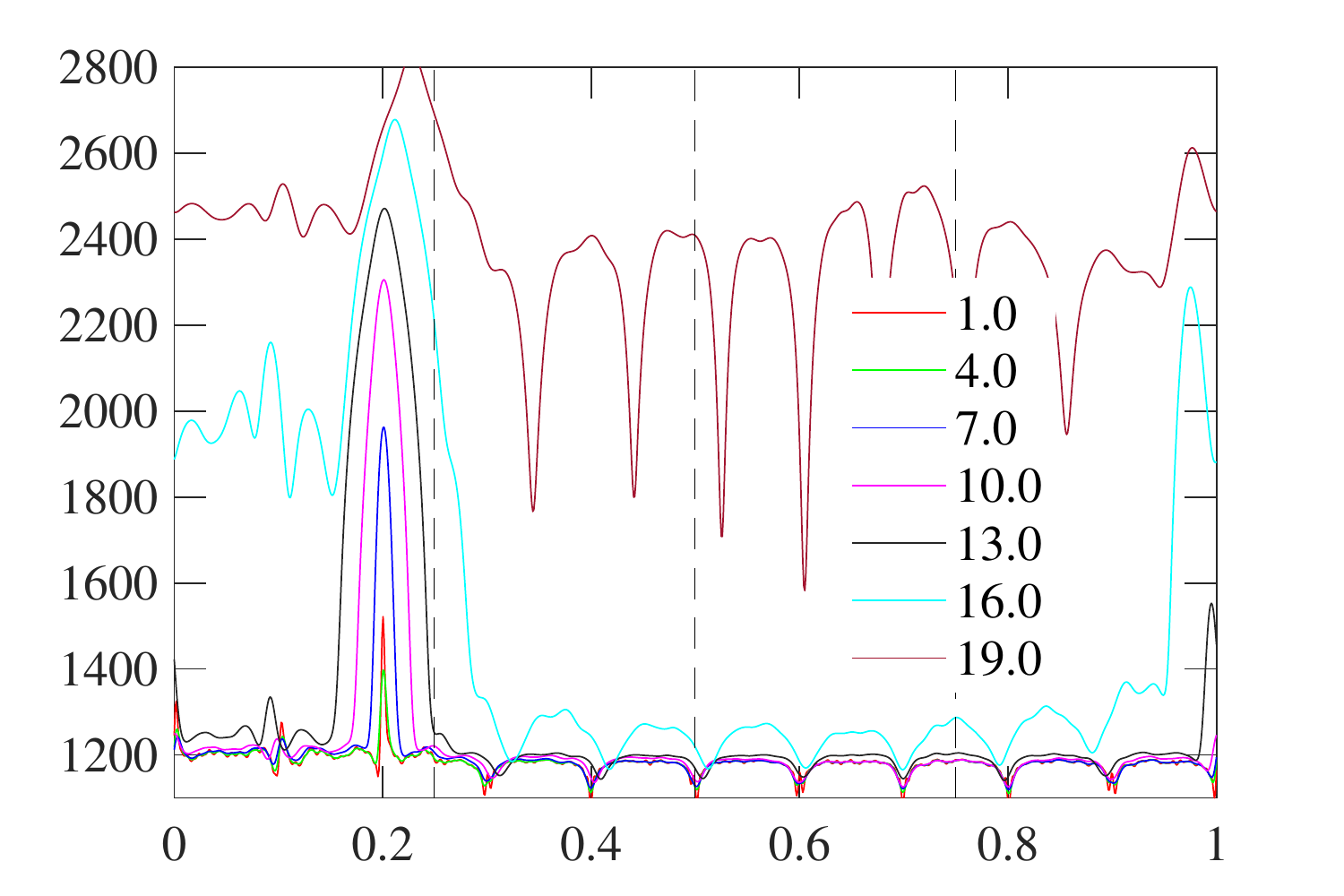}
      \bp
      \put(-160,-8){$x-position$ $(cm)$}
      \put(-265,45){\rotatebox{90}{$Temperature$ $(K)$}}
      \put(-95,120){\textbf{$Time$ $(\mu s)$}}
  \ep
  \caption{Time evolution of the temperature profiles. Dashed lines represent the sub-domain boundaries of the decomposed spatial domain.}
  \label{f:temp}
\end{figure}

The time evolution of the temperature profiles are shown in Fig. \ref{f:temp}. Initially, the temperature gradients reduce as a consequence of the diffusion process, resulting in a decrease in the peak value. The diffusion process dominates until a thermal runaway is set off due to a minor consumption of the fuel in exothermic reactions. This is first observed in the higher temperature regions which have greater propensity to trigger reactions. The thermal runaway then leads to an inception of an ignition kernel where oxidation of the fuel occurs with a rapid temperature rise and heat release. Several intermediate species are produced and consumed during the oxidation process. Eventually, regions with lower initial temperature also ignite and the combustion process completes in the entire domain.

%%%%%%%%
%%%%%%%%

To apply the anomaly detection algorithm, outlined in Algorithm 1, we decompose the spatial domain into four sub-domains ($N_d=4$), which are labeled Regions 1-4, as shown in Fig.~\ref{f:temp_ic}. In time, the dataset consists of 21 save files, which makes the number of time steps $N_t=21$. A total of 17 feature are stored in the datasets, of which 13 are the relevant features for the algorithm (12 species mass fractions and temperature). The data are pre-processed according to the scaling described in section 3.

\section{Results}
\begin{figure}[htb]
  \centering
   \includegraphics[trim=3cm 0cm 3.5cm 0cm,clip,width=3.6cm]{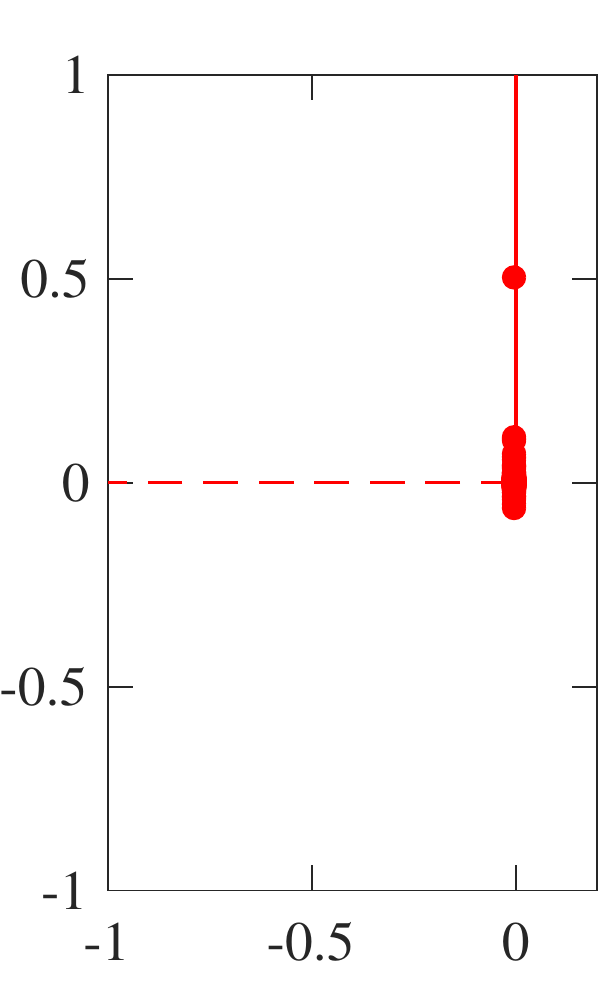}
         \bp
      \put(-75,-3){\small Mass fraction of $H_2$}
      \put(-105,60){\small \rotatebox{90}{Temperature}}
      \put(-77,135){\small (a) Time = 0 $\mu s$}
  \ep
  \hspace{0.2cm}
   \includegraphics[trim=3cm 0cm 3.5cm 0cm,clip,width=3.6cm]{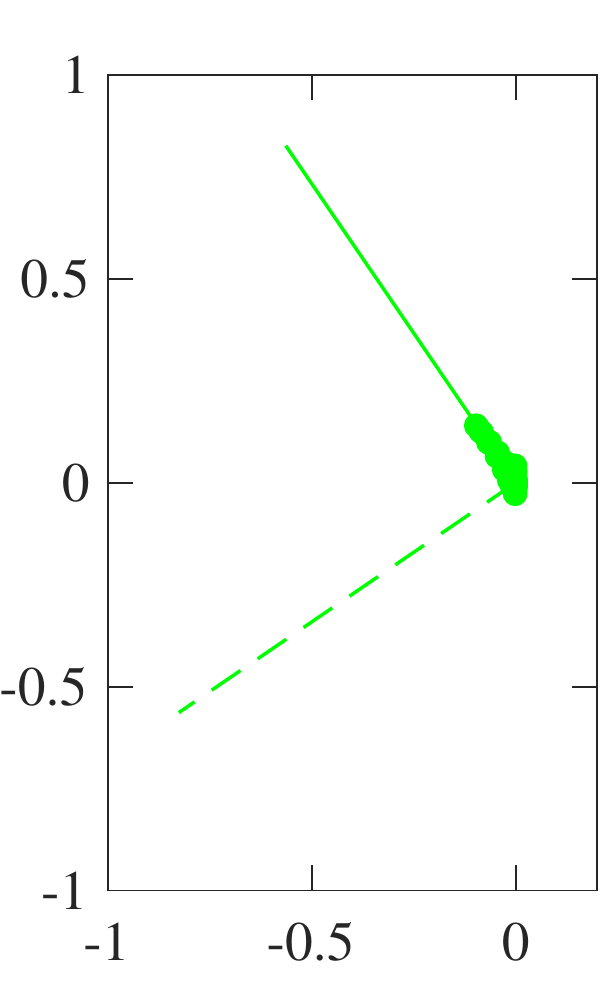}
  \bp
      \put(-75,-3){\small Mass fraction of $H_2$}
      \put(-105,60){\small \rotatebox{90}{Temperature}}
      \put(-77,135){\small (b) Time = 4 $\mu s$}
  \ep
   \hspace{0.2cm}
   \includegraphics[trim=3cm 0cm 3.5cm 0cm,clip,width=3.6cm]{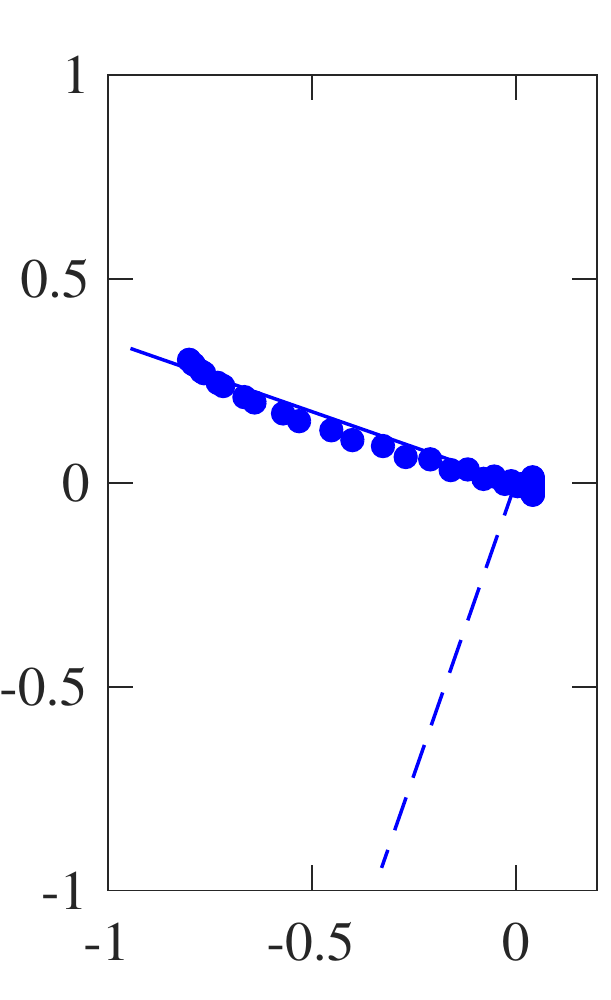}
     \bp
      \put(-75,-3){\small Mass fraction of $H_2$}
      \put(-105,60){\small \rotatebox{90}{Temperature}}
      \put(-77,135){\small (c) Time = 6 $\mu s$}
  \ep
\hspace{0.2cm}
   \includegraphics[trim=3cm 0cm 3.5cm 0cm,clip,width=3.6cm]{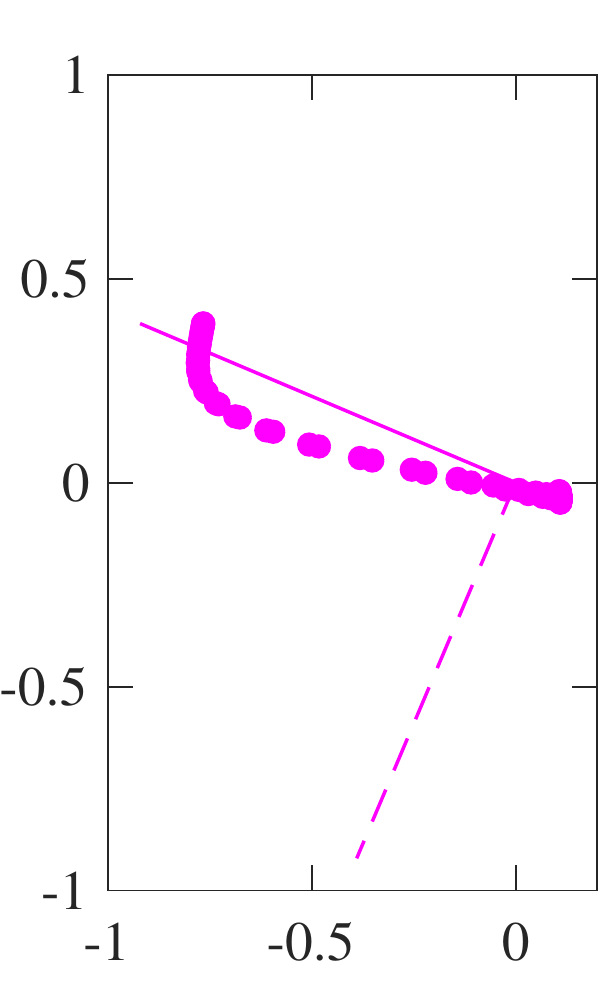}
        \bp
      \put(-75,-3){\small Mass fraction of $H_2$}
      \put(-105,60){\small \rotatebox{90}{Temperature}}
      \put(-77,135){\small (d) Time = 8 $\mu s$}
  \ep
  \caption{Scatter plots of scaled mass fraction of $H_2$ and temperature in Region 1 at different times. Solid line: first principal vector, dashed line: second principal vector.}
  \label{f:data_cloud}
\end{figure}
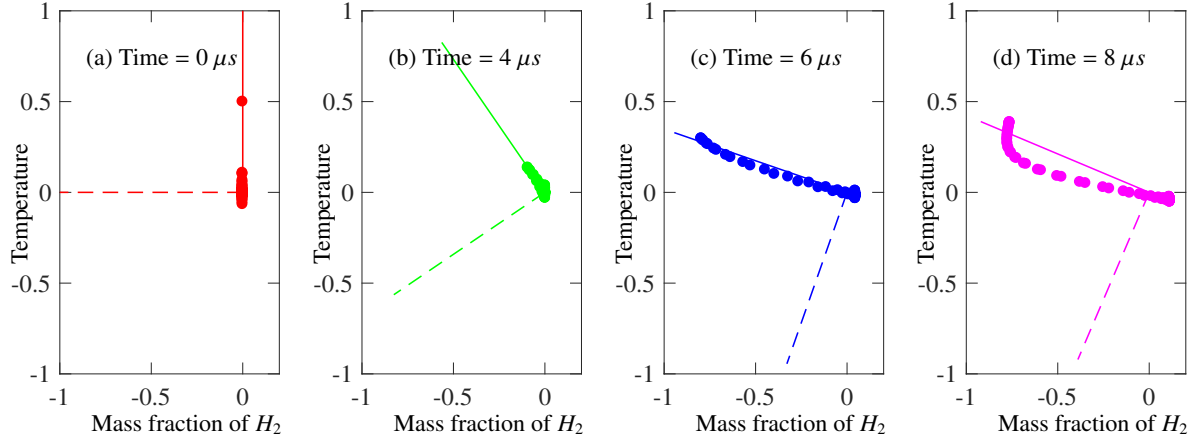

Figure~\ref{f:data_cloud} shows scatter plots of scaled mass fraction of $H_2$ and temperature for Region 1 at four different times. The solid and dashed lines in the plots represent the first and second principal kurtosis vectors, respectively. Note that for this illustration, the joint moment tensor is constructed only for these two features, which results only in two principal vectors. At initial condition the inhomogeneity in the data is present in temperature alone. Hence, the first principal vector is aligned along the temperature axis. As time advances, an ignition kernel appears in the peak temperature region, which leads to the consumption of the $H_2$-fuel and an increase in the temperature. This is depicted by some points migrating along the negative x-axis and positive y-axis. Clearly, the orientation of the principal vectors has also changed with time and the first principal vector aligns along the extreme points which represent the ignition event.

%%%%%%%%
%%%%%%%%

\begin{figure}[htb]
  \centering
  \includegraphics[width=8cm]{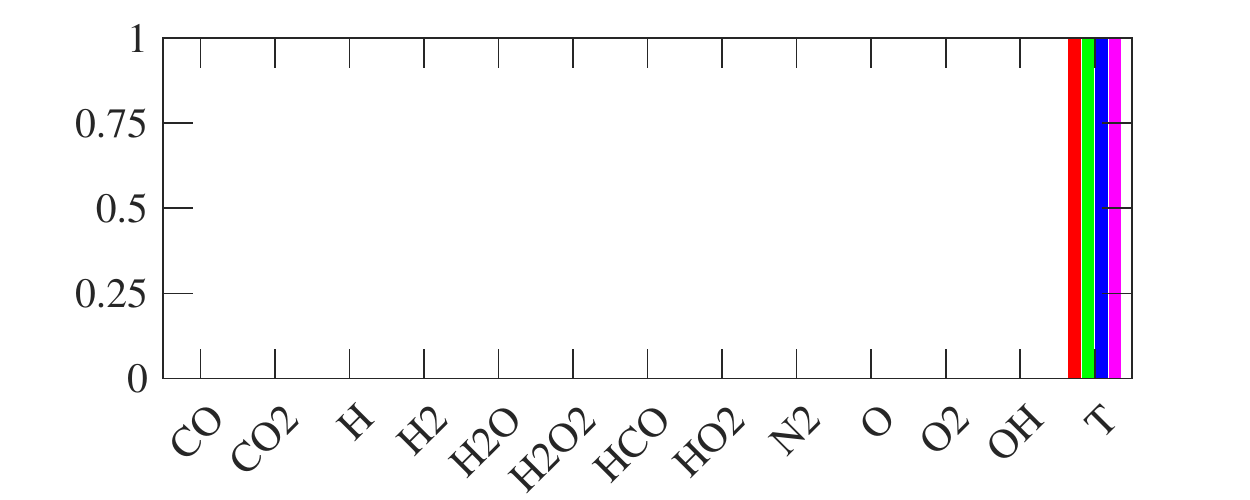}
  \bp
      \put(-230,45){\rotatebox{90}{$\F{i}{j}{n}$}}
      \put(-190,70){(a) Time = 0 $\mu s$}
      \put(-90,70){\small \colorbox{red}{\makebox(0.5,0.5){}}}
      \put(-80,69){\small Region 1}
      \put(-90,60){\small \colorbox{green}{\makebox(0.5,0.5){}}}
      \put(-80,59){\small Region 2}
      \put(-90,50){\small \colorbox{blue}{\makebox(0.5,0.5){}}}
      \put(-80,49){\small Region 3}
      \put(-90,40){\small \colorbox{magenta}{\makebox(0.5,0.5){}}}
      \put(-80,39){\small Region 4}
  \ep
  \includegraphics[width=8cm]{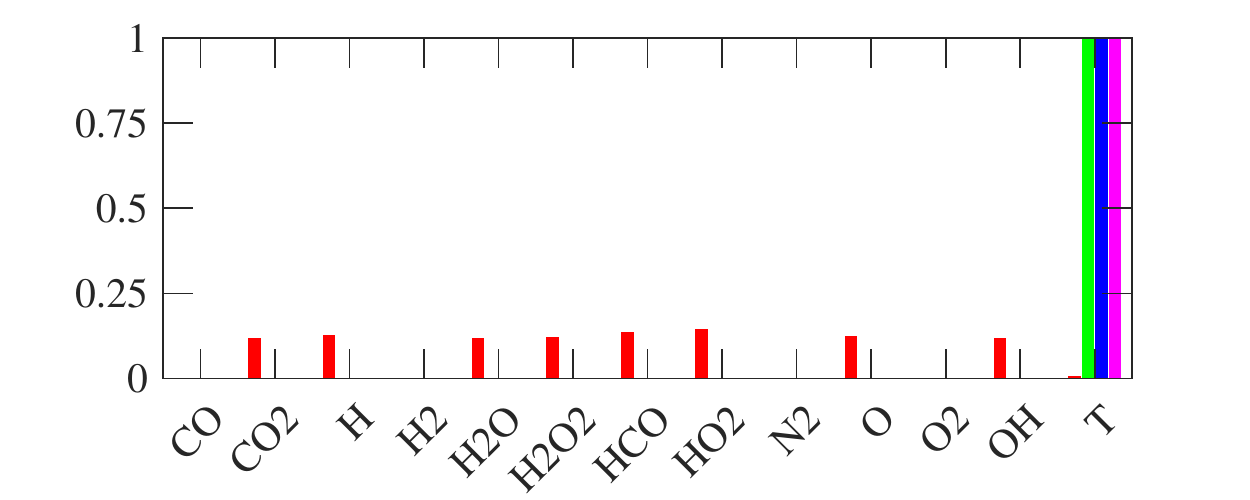}
  \bp
      \put(-230,45){\small \rotatebox{90}{$\F{i}{j}{n}$}}
      \put(-190,70){\small (b) Time = 2 $\mu s$}
  \ep
 
  \includegraphics[width=8cm]{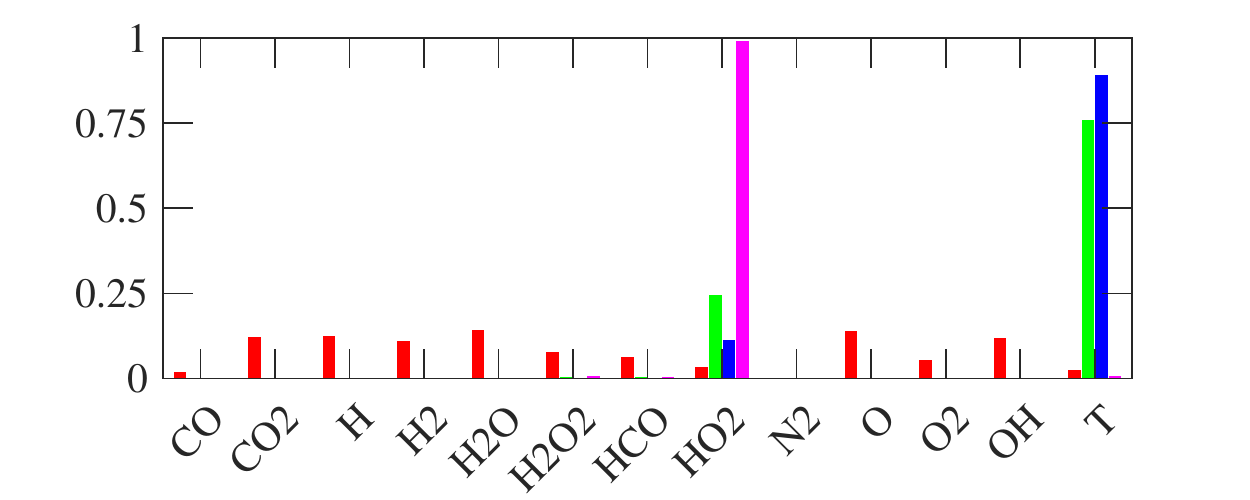}
  \bp
      \put(-230,45){\small \rotatebox{90}{$\F{i}{j}{n}$}}
      \put(-190,70){\small (c) Time = 8 $\mu s$}
  \ep
  \includegraphics[width=8cm]{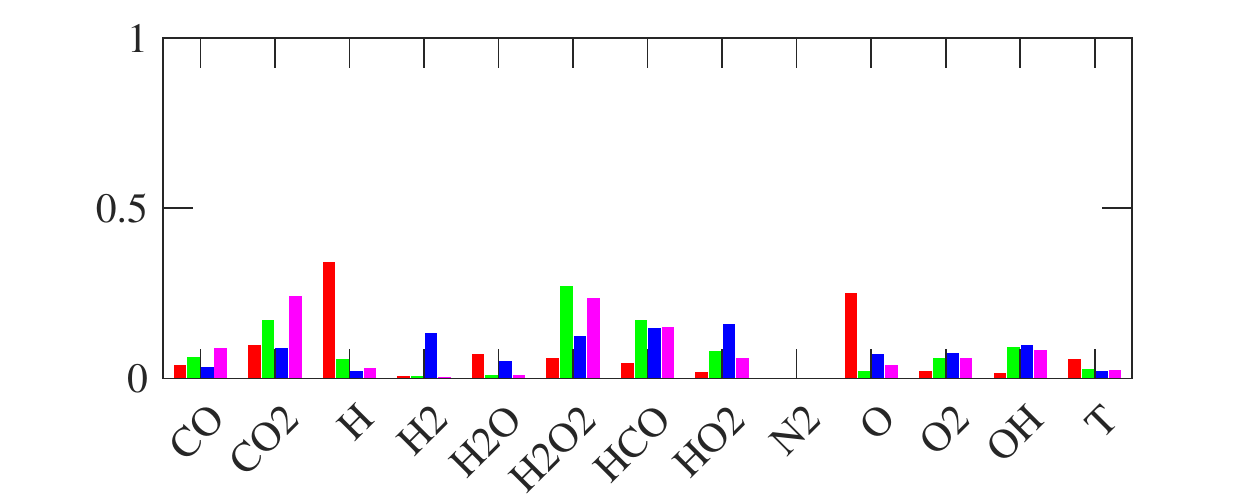}
  \bp
      \put(-230,45){\small \rotatebox{90}{$\F{i}{j}{n}$}}
      \put(-190,70){\small (d) Time = 19 $\mu s$}
  \ep
  \caption{Distribution of feature moment metric (FMM) in feature space. Different colors correspond to different sub-domains.}
  \label{f:fmm}
\end{figure}
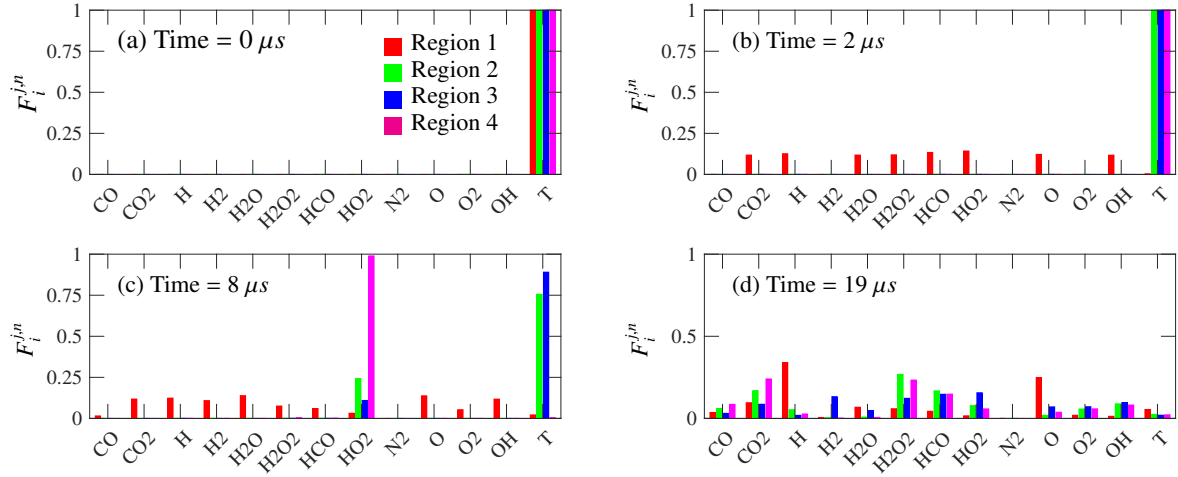

The feature moment metrics (FMMs), $\F{i}{j}{n}$, which quantify the relative importance of each feature towards the kurtosis in the data, are plotted at different times in Fig.~\ref{f:fmm}. The histograms are colored separately for the different spatial sub-domains. The x-axis consists the labels of the 13 relevant features, which include the 12 chemical species and temperature (T). At the initial time, as temperature is the only feature that consists any spread in the data in all sub-domains, the temperature-FMM is equal to 1 for all of them. As time advances to $2\mu s$ (part (b)), the FMM in temperature disappears in Region 1, and spreads to the other features, signifying a change in the FMM distribution. This is attributed to an early autoignition in the Region 1 due to the significantly greater peak temperature, which leads to locally active chemical reactions and an inhomogeneity in the mass fractions of species in the sub-domain.
At the later time, ($8\mu s$), ignition also begins in the other three sub-domains accompanied by a similar decrease in temperature-FMM. A careful look at the temperature profiles in Fig.~\ref{f:temp} will indicate that, relatively, Region 4 has the second highest temperature, followed by Region 2 and Region 3. Hence, ignition events as well as the corresponding decrease in the temperature-FMM, in the these three sub-domains will also appear in the same order in time. Eventually, ignition occurs in all the sub-domains, leading to the non-zero FMMs appearing across different features, as seen at time $19\mu s$. The FMM distributions in Fig.~\ref{f:fmm} confirm the hypothesis that at the inception of autoignition event the FMMs, which were initially significant only for temperature, begin to spread among other features, which is the change in the statistical signature anticipated. 

%%%%%%%%
%%%%%%%%

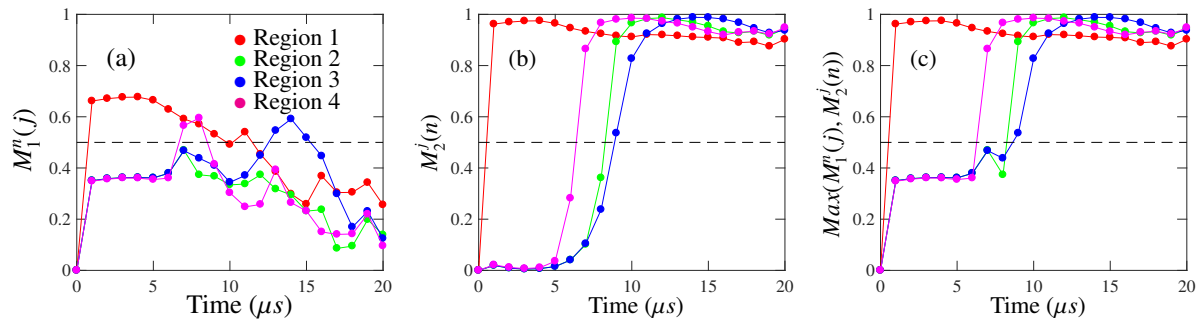
\begin{figure}[htb]
  \centering
  \includegraphics[width=5.2cm]{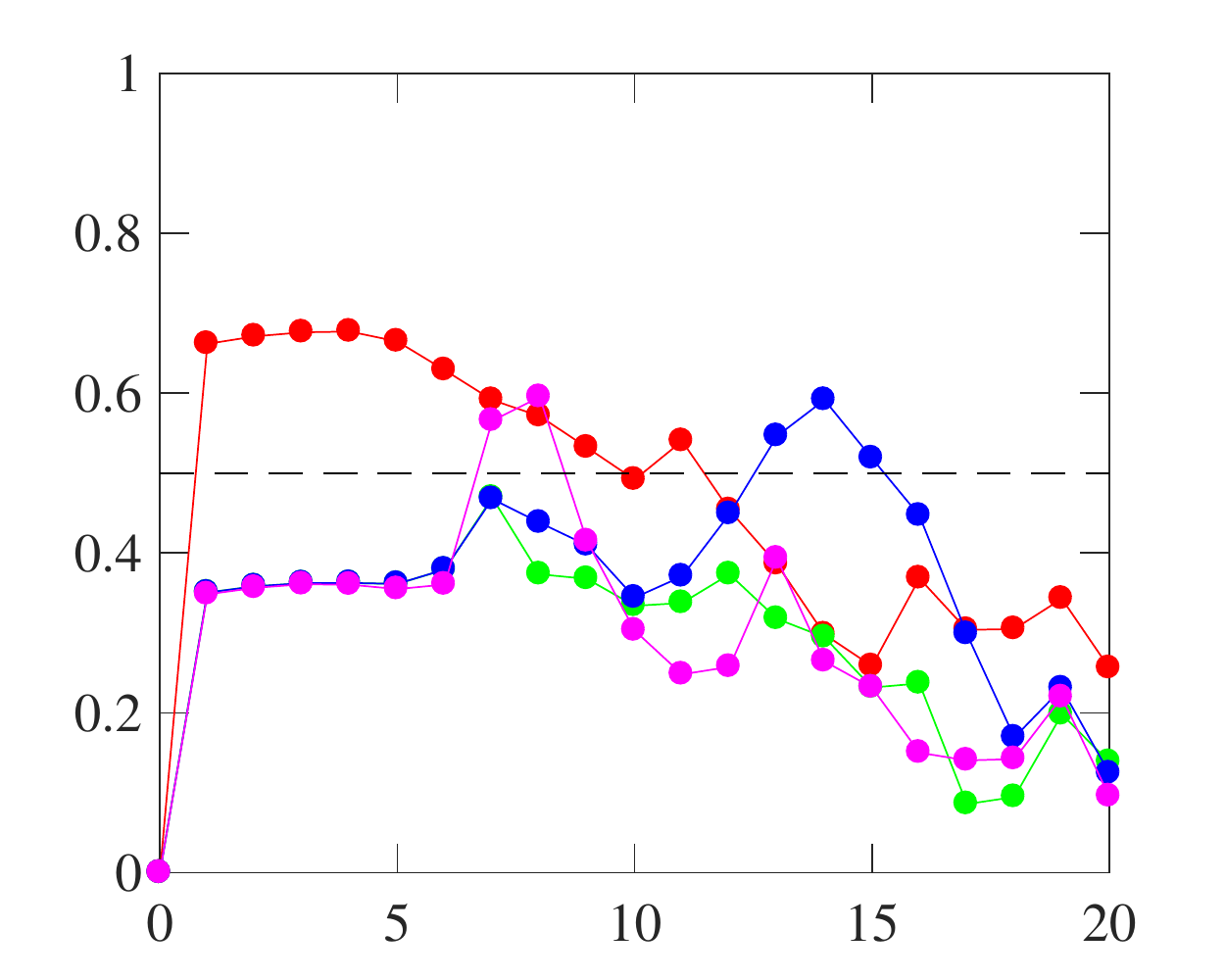}
    \bp
      \put(-155,50){\rotatebox{90}{$M_1^n(j)$}}
      \put(-90,-3){Time $(\mu s)$}
      \put(-120,90){(a)}
      \put(-70,100){\color{red} \circle*{4}}
      \put(-65,98){\small Region 1}
      \put(-70,92){\color{green} \circle*{4}}
      \put(-65,90){\small Region 2}
      \put(-70,84){\color{blue} \circle*{4}}
      \put(-65,82){\small Region 3}
      \put(-70,76){\color{magenta} \circle*{4}}
      \put(-65,74){\small Region 4}
  \ep
%   
%   \vspace{0.2cm}
  \includegraphics[width=5.2cm]{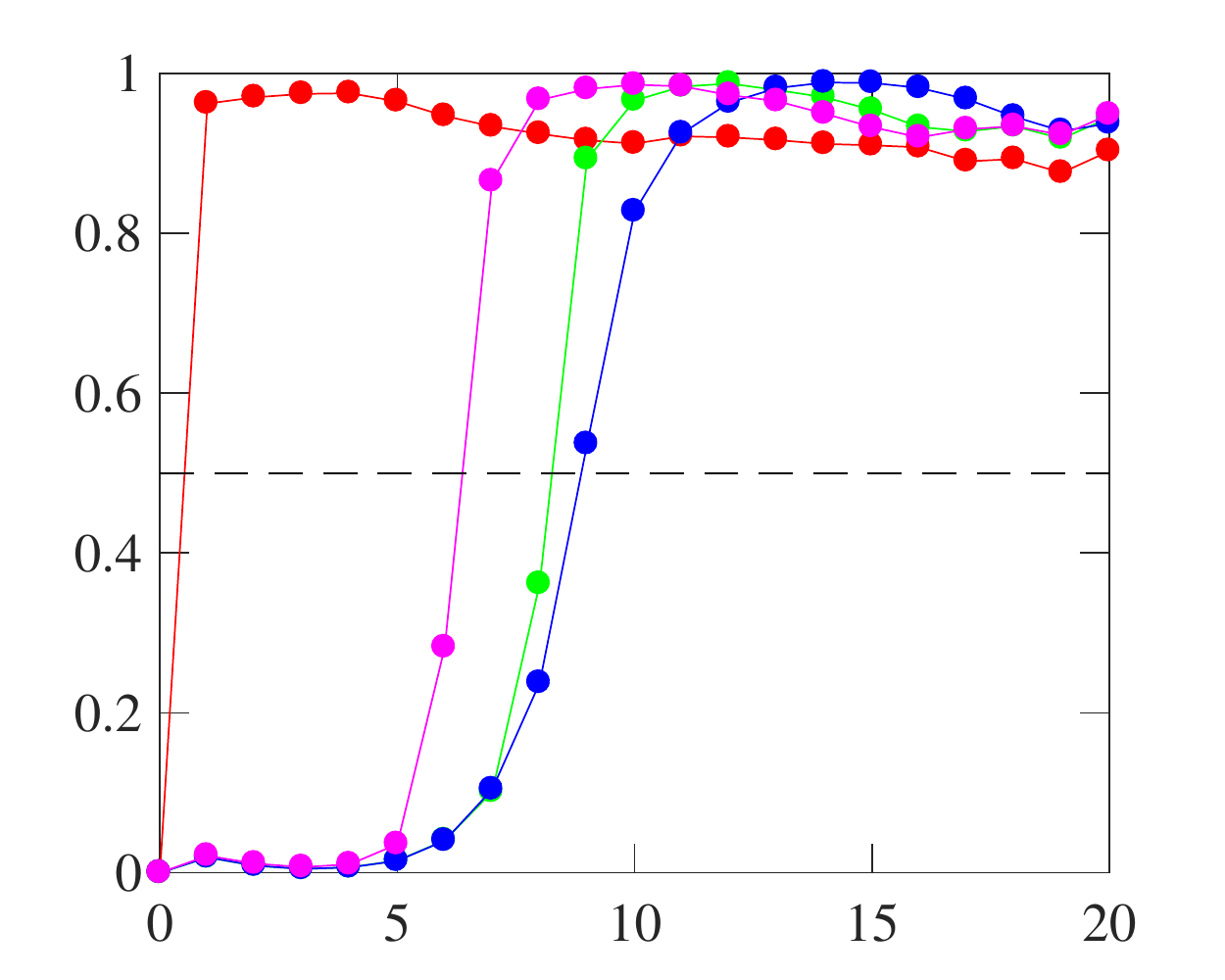}
    \bp
      \put(-155,50){\small \rotatebox{90}{$M_2^j(n)$}}
      \put(-90,-3){\small Time ($\mu s$)}
      \put(-120,90){\small (b)}
  \ep
%   
%   \vspace{0.2cm}
  \includegraphics[width=5.2cm]{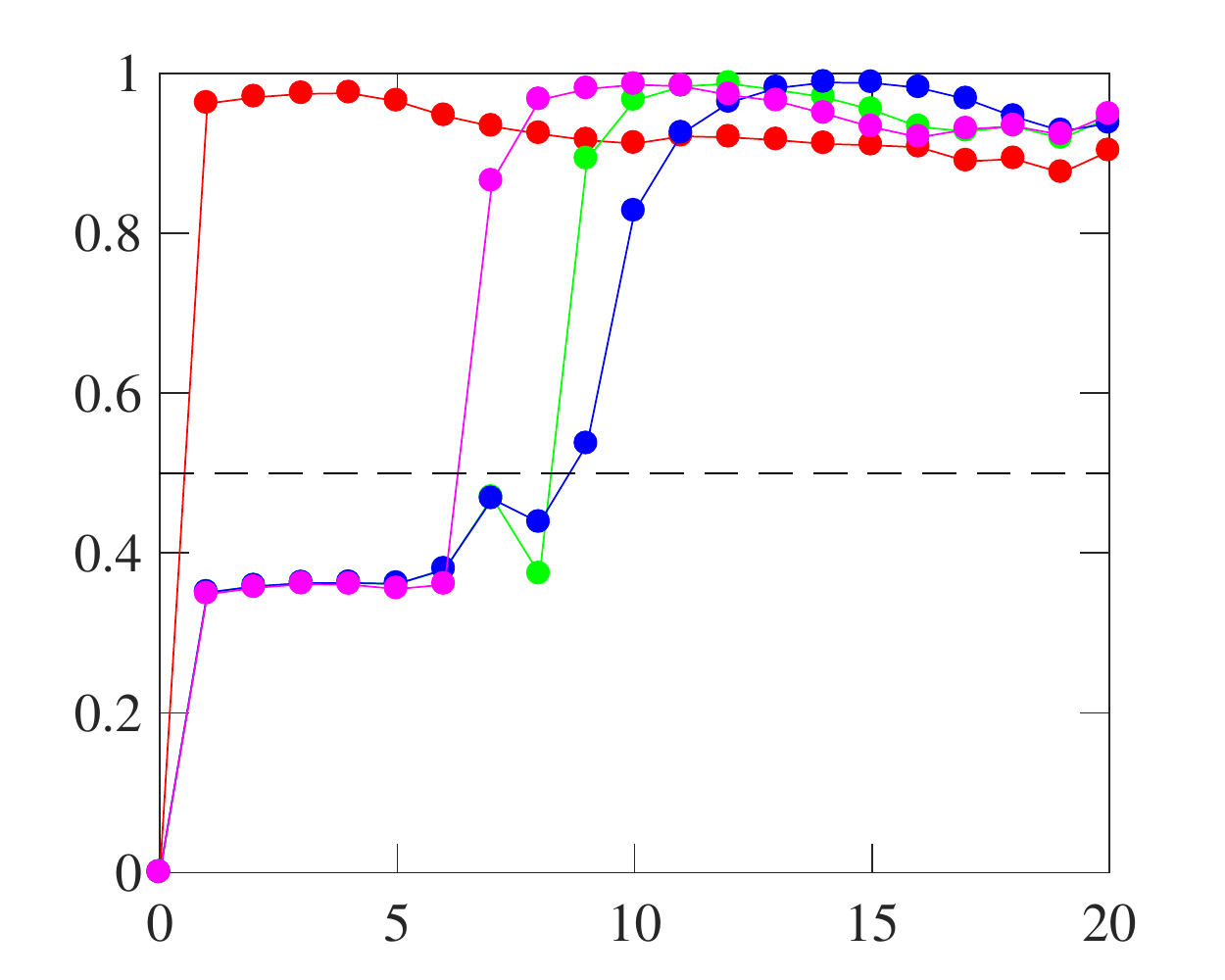}
    \bp
      \put(-155,25){\small \rotatebox{90}{$Max(M_1^n(j),M_2^j(n))$}}
      \put(-90,-3){\small Time ($\mu s$)}
      \put(-120,90){\small (c)}
  \ep
  \caption{Time evolution of (a) spatial anomaly metric, (b) temporal anomaly metric, and (c) their maximum anomaly metrics. Different colors correspond to different sub-domains. Dashed line represents a constant threshold value of 0.5.}
  \label{f:am}
\end{figure}

An anomalous autoignition event, which results in a significant change in the FMM distribution, can be detected based on the Hellinger distance, as described in section 3. The spatial and temporal anomalies are measured in terms of the distance metrics $M_1^n(j)$ and $M_2^j(n)$, respectively. Since the metrics lie between 0 and 1 we choose a threshold of $0.5$ for these metrics to identify an anomalous event. A sub-domain $j$ at a given time step $n$ can be flagged to contain an anomaly if maximum of $M_1^n(j)$ and $M_2^j(n)$ is greater than 0.5.
Fig.~\ref{f:am} shows the evolution of the $M_1^n(j)$ and $M_2^j(n)$ anomaly metrics, as well the greater of the two, with time. At earlier times ($<5\mu s$), an ignition kernel appears in the Region 1, as mentioned earlier, due to a significantly larger value of temperature. This is an anomaly both in space as well as time, because the distribution of FMMs is distinctively different from (compare parts (a) and (b) in Fig.~\ref{f:fmm}) other sub-domains and also varies quickly in time. Hence, both $M_1^n(j)$ and $M_2^j(n)$ shoot up above the threshold value. In this time period, the metrics are nearly the same and below the threshold in other sub-domains. As time progresses, beyond $5\mu s$, the temporal anomaly metric goes above the threshold in other sub-domains. For these sub-domains, the occurrence of ignition event is an anomaly in time, but not in space, and hence $M_1^n(j)$ is observed to be consistently lower than $M_2^j(n)$ in Fig~\ref{f:am}. The order in which the events appear i.e. earliest in Region 1, then in Region 4 and then in Regions 2 and 3, is also in agreement with the observed change in the FMM distribution in Fig.~\ref{f:fmm}. It appears that the metrics, and the suggested threshold, are consistent with the expected evolution of the auto-ignitive system, and are able to flag the occurrence of auto-ignition in the right sub-domains and at the right times.

%The results presented in this section demonstrate the robustness of the proposed algorithm in detecting spatial and/or temporal anomalies.

%%%%%%%%
%%%%%%%%

\section{Conclusions}
%Motivation, approach idea, background, algorithm, metrics, test case, results, applications, future work.

Detection of anomalous events in scientific phenomena remains a key challenge, particularly, due to the availability of increasingly high resolution data which is enabled by the advances in measurement techniques and computing power.
Scientific datasets are characterized by smoothly varying multi-variate data which represent complex non-linear multi-physics processes.  
Hence, commonly used anomaly detection algorithms developed for the use in other domains may not be readily applicable for scientific data. 
It is well known that occurrence of anomalies manifest as extreme values in the distribution of at least some of the features, and significantly affect their higher order joint statistical moments. In this paper, we have used this idea to develop a robust anomaly detection algorithm for scientific data.

For normally distributed data, the statistical information is fully encapsulated in the second order co-variance matrix. However, scientific data are often non-Gaussian and need further higher order joint moment tensors to characterize the anomalous events. By analogy to principal component analysis, a general joint moment tensor can be analyzed in terms of its principal values and vectors which can be computed using symmetric tensor decomposition methods. We have reviewed commonly used methods and identified that a simple singular value decomposition of the matricized tensor works best for our purpose. Using synthetic data, we have shown that vectors computed from the decomposition of cumulant fourth moment tensor, also referred to as excess-kurtosis, capture the outlier data perfectly.

In general, anomalous events in scientific phenomena appear in space and/or time. To identify the locality of an anomaly, the proposed algorithm decomposes the data into spatial sub-domains and time steps. Feature moment metrics are computed for each sub-domain and time step to quantify the magnitude and alignment of principal values and vectors, respectively. With an inception of an anomaly the distribution feature moment metrics significantly change. The algorithm, then, uses Hellinger distance to compare the feature moment metrics between different sub-domains and successive time steps to flag anomalies in space and time, respectively. The algorithm has been tested using a turbulent combustion test case to detect autoignition events.

The unsupervised anomaly detection algorithm presented in this paper has been shown to robustly identify the spatial and temporal anomalies. The statistical approach in the algorithm uses a decomposed layout of the data which is inherent to large streaming distributed scientific datasets. Our future work includes two aspects. First, an in-situ implementation of the algorithm into the massively parallel direct numerical simulation solver (S3D) and evaluate its scalability. Second, to apply the algorithm to detect anomalies in other scientific phenomena. \\

%%%%%%%%
\section{Acknowledgments}
This work was funded through U.S. Department of Energy Advanced Scientific Computing Research (ASCR) grant FWP16-019471. Sandia National Laboratories is a multi-mission laboratory managed and operated by National Technology and Engineering Solutions of Sandia, LLC.,a wholly owned subsidiary of Honeywell International, Inc., for the U.S. Department of Energy’s National Nuclear Security Administration under contract DE-NA-0003525. The views expressed in the article do not necessarily represent the views of the U.S. Department of Energy or the United States Government.

% \appendix
% \section{}
% \label{}

% References
%
% Following citation commands can be used in the body text:
% Usage of \cite is as follows:
%   \cite{key}          ==>>  [#]
%   \cite[chap. 2]{key} ==>>  [#, chap. 2]
%   \citet{key}         ==>>  Author [#]

% References with bibTeX database:

\bibliographystyle{model1-num-names}
\bibliography{main.bib}

% Authors are advised to submit their bibtex database files. They are
% requested to list a bibtex style file in the manuscript if they do
% not want to use model1-num-names.bst.

% References without bibTeX database:


% \bibitem must have the following form:
%   \bibitem{key}...
%


% \end{thebibliography}

\end{document}